\documentclass[pre,amsmath, twocolumn, floatfix,showpacs]{revtex4}
\usepackage{graphicx}
\usepackage{amssymb}
\usepackage{amsmath}

\begin{document}
\title{Comment on ``A generalized Langevin formalism of complete DNA
melting transition"}
\author{Titus S. van Erp$^{1}$, Santiago Cuesta-Lopez$^{2}$,
Johannes-Geert
Hagmann$^{2}$, and Michel Peyrard$^2$}

\affiliation{
 1: Centrum voor Oppervlaktechemie
en Katalyse, K.U. Leuven, Kasteelpark Arenberg 23, B-3001 Leuven,
Belgium\\
2: 
 Laboratoire de Physique, Ecole Normale Sup\'erieure de Lyon, 
46 all\'ee d'Italie, 69364 Lyon Cedex 07, France
} 

\begin{abstract}
We show that the calculated DNA denaturation curves for
finite (Peyrard-Bishop-Dauxois (PBD) chains are intrinsically undefined.
\end{abstract}


\maketitle

In Ref. \cite{Das}, the authors claim to solve a problem of 
{\em  equilibrium}
statistical physics by increasing damping in a Langevin simulation, i.e.\ by
changing the {\em dynamics}. 
This is a fundamental flaw which though forms the basis of the
work. Increasing $\Gamma$ 
may change how fast the phase space is explored, but, if the simulations are
performed properly, it cannot alter the equilibrium distribution.
The modification introduced by the authors apparently solves the
problem of the divergence of $\langle y \rangle$ as the simulations 
start from a closed initial condition instead of using a set of initial 
conditions
distributed over the whole phase space. 
The large increase of damping
considerably slows down the exploration of phase space in the calculation,
so much that the system stays in the vicinity of the closed
initial configuration.

Moreover, the dynamical argument used by the authors does not stand
examination
because it underestimates the actual time-scale which is 
provided 
by the experiments. 
Using the standard diffusion equation (with corrected prefactor compared to
\cite{Das}) 
\begin{equation}
P(y,t;y_0)=\big(\frac{m \Gamma}{4 \pi k_B T t}\big)^{1/2} \exp(-m \Gamma
(y-y_0)^2/4 k_B T t), 
\end{equation}  
the authors state that \emph{''the probability of attaining infinitely large
  values of the bead displacements is effectively zero''}.   
Although strictly speaking correct, it is easy to show that this does not
prevent divergence on the experimental timescale, which  is typically  2
min. per temperature step.  
Using the variables $\Gamma= 0.5~ {\rm ps}^{-1}$, ${\rm m}=350~ {\rm u}$, and
$T=300~ {\rm K}$, the chance that the two beads are within reasonable
distance, e.g $<100$ \AA~,  
is still large when the timescales of computer simulations are considered,
i.e.\ 
$\approx  94 \%$ for $t=1~ {\rm ns}$. However, taking the experimentally
relevant scale of $t=2$ min., this probability has dropped to less than 
$2 \cdot 10^{-6}$. 

A Monte Carlo study of the same system\cite{Ares} 
also suffers from this problem as we showed in a recent paper\cite{us}.  
The large quantitative differences between ~\cite{Das} and \cite{Ares} should
have alarmed 
the authors that at least one of these studies 
reports unconverged results. 
Instead, 
the results reported in~\cite{Das}  are found to be in remarkable agreement 
with the former study~\cite{Ares} and
experimental results~\cite{Montri}. 
In principle,  both Langevin~\cite{Das} and MC~\cite{Ares} 
should unsurprisingly sample the same
distribution.  
Yet, although the curves look qualitatively similar,  
for some temperatures they show values that differ by more than $40 \%$.  
As a general trend, when compared to experiments~\cite{Montri}
the curves are shifted towards lower values
along the  temperature axis. 
A recent comment~\cite{Joy} on~\cite{Ares}, showed that denaturation curves,
calculated in this way,  continue to shift to 
lower temperatures when the simulation time is extended. 
The apparent signature of premelting in ~\cite{Das,Ares}
that has been observed experimentally~\cite{Montri}, 
is actually a numerical artifact in the theoretical studies. The present PBD 
model is unable to reproduce this experimental feature as can 
be concluded from more accurate numerical studies~\cite{us,Joy}.

The misconceptions appearing in this work and others on the same subject show
that it is necessary to clarify some essential principles regarding the
modeling of thermal DNA denaturation and on the computational methods used to
address the problem. 
The Peyrard-Bishop-Dauxois (PBD)\cite{PBD} model has
some fundamental limitations which seem to be 
ignored~\cite{us}.
It was not designed to sample the phase space corresponding to fully melted
DNA molecules. Therefore, computing experimental quantities such as $p$
(fraction of fully melted molecules) does not make sense for several
reasons.

First, as discussed above, the phase space that comprises the fully
denaturated molecules is infinitely larger than that of the (partially) closed
molecules. As result, the DNA  molecule will always  
visit and reside in
the fully denaturated state at any temperature. The reason that this 
does not appear in the experiments is because there is an
equilibrium between melting and self assembly in DNA solutions.
Conversely, the PBD model describes a single DNA molecule in an infinitely
diluted system. 

Secondly,  each base is linked to its complementary base by a given on-site
  potential, determined by the initial state. Therefore assembly /disassembly
  reactions are unphysically favored in a particular pathway.
Equilibrium intermediate states are not well sampled as mismatches are not
  considered, while these structural events play a very important role in the
  premelting phase, specially in bubble nucleation and dynamics.

In spite of its limitations the
model can  give useful results in the
temperature range where the molecule is not fully denaturated (dsDNA
ensemble \cite{us}). Surprisingly,
the denaturation of infinite DNA chains can also be properly described. 
The presence of this other infinity, besides that of the unbounded phase space,
ensures that the molecule remains in the double stranded 
state at all times~\cite{us}. This can be understood via the following 
reasoning: when moving in one
direction along the chain, the sequence $y_i,y_{i+1},\ldots$ basically 
represents an infinite one-dimensional random walk which must 
always visit the origin (assigning the whole molecule as being 'closed') 
according to P\'olya's theorem.

Moreover, recent 
technological progress~\cite{Montri}  
makes it  possible to isolate
the experimental signal of  partially closed molecules from the fully
denaturated population.  For this 
reason we introduced the bias potential $V_{\rm bias}(y_{\rm min})$ which
makes it possible  
to study the PBD model at elevated  temperatures. 
In particular, we focused on the $l-$denaturation curves which describe the
fraction of open base-pairs 
as function of temperature within the population of closed molecules. This
property is well  
defined and gives a more direct comparison with experiments~\cite{Montri} than
the  phenomenological equation taking
into account the  fully open molecules\cite{Campa}, which 
requires additional fitting parameters. They are hard to 
determine properly
because they appear in exponential terms and the results are very
sensitive to their values.

The criticism on this  bias potential $V_{\rm bias}(y_{\rm min})$ is 
unfounded.
The article claims that there is no rational  justification of this particular
form, but, 
in fact, $V_{\rm bias}$   can be any kind of steep increasing function. The
exact shape has no influence on the results as it only affects the tail
of the  
distribution. Our results were actually not carried out by MD or MC, but using
a much more efficient  
transfer integral approach~\cite{us2} for which  $V_{\rm bias}$ is simply an
infinite step-function.  
Also the second argument that it lowers the equilibrium values of $f$  (
fraction of open base pairs considering all molecules), $p$ and $l$ is
untrue. $f$  and $p$ are ill-defined anyway and therefore we did not report on
them. However, $l(T)$ is perfectly defined and remains exact when the bias
potential is invoked. 
As  $V_{\rm bias}(y_{\rm min})$ only excludes parts of phase space that does
not contribute to $l$, it has 
no effect on it.  
This can easily be verified mathematically but has also been confirmed by
unbiased  MC simulations~\cite{Joy}.

To conclude,
the simulation approach applied in this work~\cite{Das} and 
other 
related studies, e.g.\ ~\cite{Ares}, is not well founded in terms of
statistical physics and uses the PBD model in situations where it can not be  
expected to be valid. 
The apparent signatures of premelting in ~\cite{Das,Ares} 
 are nothing more than a memory effect of chosen initial conditions and 
 do not hold when they are scrutinized by more accurate
 calculations~\cite{us,Joy}.   

\acknowledgments
TSvE acknowledge the Flemish government for a concerted
research action (GOA) and the Flemish FWO for financial support.

\bibliographystyle{prsty}

\end{document}